\documentclass[prd,preprint,tightenlines,superscriptaddress,showpacs,byrevtex]{revtex4}

\usepackage{graphicx} 
\usepackage{dcolumn}  
\usepackage{bm}

\graphicspath{{ps}}

\begin{document}

\begin{minipage}[t]{3cm}
\includegraphics[width=3cm]{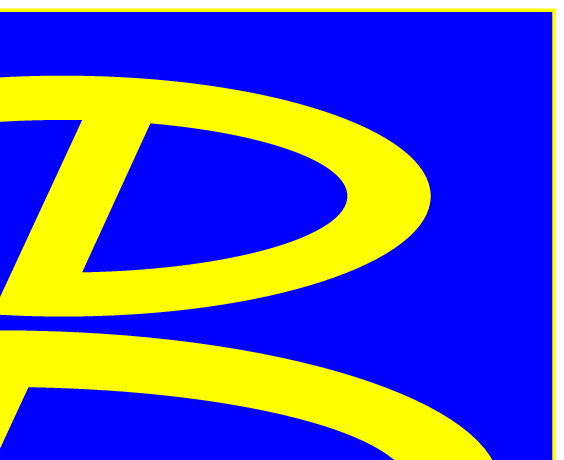}
\end{minipage}
\begin{minipage}[t]{16cm}
\hspace*{8cm}{KEK Preprint 2002-96}\\
\hspace*{8cm}{Belle Preprint 2002-32}
\end{minipage}

\title{
 \quad\\[1cm] \Large
Measurement of Branching Fractions and Charge Asymmetries \\
for Two-Body $B$ Meson Decays with Charmonium}


\affiliation{Budker Institute of Nuclear Physics, Novosibirsk}
\affiliation{Chuo University, Tokyo}
\affiliation{University of Cincinnati, Cincinnati, Ohio 45221}
\affiliation{Gyeongsang National University, Chinju}
\affiliation{University of Hawaii, Honolulu, Hawaii 96822}
\affiliation{High Energy Accelerator Research Organization (KEK), Tsukuba}
\affiliation{Institute of High Energy Physics, Chinese Academy of Sciences, Beijing}
\affiliation{Institute of High Energy Physics, Vienna}
\affiliation{Institute for Theoretical and Experimental Physics, Moscow}
\affiliation{J. Stefan Institute, Ljubljana}
\affiliation{Kanagawa University, Yokohama}
\affiliation{Korea University, Seoul}
\affiliation{Kyoto University, Kyoto}
\affiliation{Kyungpook National University, Taegu}
\affiliation{Institut de Physique des Hautes \'Energies, Universit\'e de Lausanne, Lausanne}
\affiliation{University of Ljubljana, Ljubljana}
\affiliation{University of Maribor, Maribor}
\affiliation{University of Melbourne, Victoria}
\affiliation{Nagoya University, Nagoya}
\affiliation{Nara Women's University, Nara}
\affiliation{National Lien-Ho Institute of Technology, Miao Li}
\affiliation{National Taiwan University, Taipei}
\affiliation{H. Niewodniczanski Institute of Nuclear Physics, Krakow}
\affiliation{Nihon Dental College, Niigata}
\affiliation{Niigata University, Niigata}
\affiliation{Osaka City University, Osaka}
\affiliation{Osaka University, Osaka}
\affiliation{Panjab University, Chandigarh}
\affiliation{Peking University, Beijing}
\affiliation{Princeton University, Princeton, New Jersey 08545}
\affiliation{Saga University, Saga}
\affiliation{University of Science and Technology of China, Hefei}
\affiliation{Seoul National University, Seoul}
\affiliation{Sungkyunkwan University, Suwon}
\affiliation{University of Sydney, Sydney NSW}
\affiliation{Tata Institute of Fundamental Research, Bombay}
\affiliation{Toho University, Funabashi}
\affiliation{Tohoku Gakuin University, Tagajo}
\affiliation{Tohoku University, Sendai}
\affiliation{University of Tokyo, Tokyo}
\affiliation{Tokyo Institute of Technology, Tokyo}
\affiliation{Tokyo Metropolitan University, Tokyo}
\affiliation{Tokyo University of Agriculture and Technology, Tokyo}
\affiliation{Toyama National College of Maritime Technology, Toyama}
\affiliation{University of Tsukuba, Tsukuba}
\affiliation{Utkal University, Bhubaneswer}
\affiliation{Virginia Polytechnic Institute and State University, Blacksburg, Virginia 24061}
\affiliation{Yokkaichi University, Yokkaichi}
\affiliation{Yonsei University, Seoul}
  \author{K.~Abe}\affiliation{High Energy Accelerator Research Organization (KEK), Tsukuba} 
  \author{R.~Abe}\affiliation{Niigata University, Niigata} 
  \author{T.~Abe}\affiliation{Tohoku University, Sendai} 
  \author{I.~Adachi}\affiliation{High Energy Accelerator Research Organization (KEK), Tsukuba} 
  \author{H.~Aihara}\affiliation{University of Tokyo, Tokyo} 
  \author{M.~Akatsu}\affiliation{Nagoya University, Nagoya} 
  \author{Y.~Asano}\affiliation{University of Tsukuba, Tsukuba} 
  \author{T.~Aso}\affiliation{Toyama National College of Maritime Technology, Toyama} 
  \author{T.~Aushev}\affiliation{Institute for Theoretical and Experimental Physics, Moscow} 
  \author{A.~M.~Bakich}\affiliation{University of Sydney, Sydney NSW} 
% ...
\collaboration{The Belle Collaboration}

\begin{abstract}

We report branching fractions and charge asymmetries for exclusive
decays of charged and neutral $B$ mesons to two-body final states
containing a charmonium meson, $J/\psi$ or $\psi$(2S).  This result is
based on a $29.4~{\rm fb}^{-1}$ data sample collected at the
$\Upsilon$(4S) resonance with the Belle detector at the KEKB
asymmetric $e^+e^-$ collider.
\end{abstract}


\maketitle

\clearpage

\section{Introduction}

Investigation of $CP$ violation is one of the key issues facing
elementary particle physics. 
Recently, BaBar\cite{BabarCPV} and Belle\cite{BelleCPV}
have observed large time-dependent $CP$ asymmetries 
in the neutral $B$-meson system\cite{CPV}.
Decay modes of neutral $B$ mesons to final states containing
charmonia were used for these measurements due to their clean
experimental signatures and straightforward theoretical
interpretation.
It is expected for the same reasons that exclusive charmonium
modes will continue to play a major role in $CP$ studies, with 
rarer modes contributing as the body of data grows in magnitude
and different aspects of the $CP$ question move to the forefront.
For example, the Kobayashi-Maskawa model\cite{KM} predicts small
direct $CP$ violation for $B\rightarrow J/\psi K^{\pm}$ and $B\rightarrow
J/\psi\pi^{\pm}$\cite{DCPV_jpsih}.
Large direct $CP$ violation would indicate new physics\cite{DCPV}.
In addition, the dominant mechanism for charmonium production
in $B$-meson decay is color suppressed, so precise measurements
of rates to the exclusive modes can provide important information
toward the understanding of color suppression.

In this paper we report measurements of branching fractions and
charge asymmetries for the exclusive decays of $B$ mesons to the
two-body final states $\psi h$, where $\psi$ is $J/\psi$ or $\psi$(2S)
and $h$ is one of the light mesons $K^{\pm}$, $K^0_S$, $\pi^{\pm}$, or
$\pi^0$. We used a $29.4~{\rm fb}^{-1}$ data set which contains 31.9 million
$B\bar{B}$ events collected with the Belle detector\cite{Belle} at
KEKB\cite{KEKB}.

\section{The Belle Detector}

KEKB is an asymmetric electron-positron storage ring that collides
8.0 GeV electrons with 3.5 GeV positrons at the $\Upsilon$(4S)
resonance (10.58 GeV center-of-mass energy). The $\Upsilon$(4S)
resonance is boosted by $\beta \gamma = 0.425$. There is a 22~mrad
crossing angle between the electron and positron beams at the interaction
point.

The Belle detector surrounds the beam crossing point.  It is a large
solid angle spectrometer with a 1.5 T superconducting solenoid magnet.
Charged particles are detected by a three layer double-sided silicon
vertex detector (SVD) and a 50 layer cylindrical drift chamber (CDC)
filled with a helium-ethane gas mixture.
The tracking acceptance covers the laboratory polar angle between 
$\theta = 17^\circ$ and $150^\circ$ ($z$ is along the beam direction),
corresponding approximately to
92\% of the full solid angle in the center-of-mass frame (CM).
The resolutions in impact parameter and momentum are measured to be 
55 $\mu$m for a 1 GeV/$c$ charged particle and 
$\sigma_{p_t}/p_t = (0.30/\beta \oplus 0.19p_t)$\%, where $p_t$ is the
transverse momentum in GeV/$c$.
A CsI(Tl) electromagnetic calorimeter (ECL) is located inside the solenoid coil
and covers the same solid angle as the charged particle tracking system. It 
detects electromagnetic showers with a resolution of 
$\sigma_E/E = (1.3\oplus0.07/E\oplus0.8/E^{1/4})$\%, where $E$ is in GeV.

Charged hadron identification is accomplished by combining 
the response from an array of 1188 silica aerogel 
\v{C}erenkov counters (ACC), an array of 128 time-of-flight counters
(TOF) and specific ionization ($dE/dx$) measurement in the CDC.
An iron flux-return yoke outside the solenoid is comprised of 
14 layers of 4.7 cm-thick iron plates interleaved with a system of 
resistive plate counters (KLM) that are used for muon identification.
The Belle detector is described in detail elsewhere\cite{Belle}.
\section{Event selection}

Hadronic events are selected by requiring 
(1) at least three reconstructed charged tracks, 
(2) a total reconstructed ECL energy in the
CM in the range 0.1 to 0.8 times the total CM energy, 
(3) at least one large-angle cluster in the ECL,
(4) a total visable energy (sum of charged tracks and neutral showers
not matched to tracks)
greater than 0.2 times the total CM energy,
(5) absolute value of the $z$ component of the CM momentum less than
50\% of the total CM energy, and
(6) a reconstructed primary vertex that is consistent with the known
location of the interaction point.
These selection criteria are determined by Monte Carlo simulation to
be 99\% efficient for signal events.
To suppress two-jet non-$\Upsilon$(4S) background
relative to $B\bar{B}$ events we require
that $R_2<0.5$, where $R_2$ is the ratio of the second to zeroth
Fox-Wolfram moments\cite{Fox-Wolfram}.
To remove charged particle tracks that are badly measured or do not
come from the interaction region, we require $dz < 5$~cm for all
tracks other than those identified as decay daughters of $K^0_S$,
where $dz$ is the absolute value of the coordinate along the beam
direction at the point on the track nearest the origin.

The decay modes considered are listed in Table~\ref{decay_modes}. (Hereafter
the inclusion of the charge conjugate states is implied.)

\begin{table}[htb]
\caption{\label{decay_modes}Analyzed decay chains.}
\begin{ruledtabular}
\begin{tabular}{ll}
 Primary  mode & Secondary mode(s)       \\
\hline
$B^-       \to J/\psi\pi^-     $ & $J/\psi \to l^+l^- $\\ 
$\bar{B^0} \to J/\psi\pi^0     $ & $J/\psi \to l^+l^- $, $\pi^0 \to \gamma\gamma $ \\
$B^-       \to J/\psi K^-      $ & $J/\psi \to l^+l^- $ \\
$\bar{B^0} \to J/\psi K^0_S    $ & $J/\psi \to l^+l^- , K^0_S \to \pi^+\pi^-$\\
$B^-       \to \psi{\rm(2S)}K^-$ & $\psi {\rm(2S)} \to l^+l^-$, 
$\psi {\rm(2S)} \to J/\psi\pi^+\pi^-\{J/\psi \to l^+l^- \}$\\
$\bar{B^0} \to \psi{\rm(2S)}K^0_S$ & $\psi {\rm(2S)} \to l^+l^-$, 
$\psi {\rm(2S)} \to J/\psi\pi^+\pi^-\{J/\psi \to l^+l^- \},\ K^0_S \to \pi^+\pi^-$\\
\end{tabular}
\end{ruledtabular}
\end{table}

\subsection{$J/\psi$($\psi$(2S)) Candidates}

In this analysis, $J/\psi$ candidates are reconstructed from oppositely
charged lepton pairs, $\mu^+\mu^-$ or $e^+e^-$.  Lepton candidates are
selected with tight or loose criteria depending on the background
level for each mode. For muon tracks, tight identification is based
on track penetration depth and hit scatter in the
KLM system, while loose identification requires that the track
have an energy deposit in the ECL that is consistent with that of a
minimum ionizing particle. Electron tracks are tightly identified
by a combination of $dE/dx$ from the CDC, $E/p$ ($E$ is energy deposit
in the ECL and $p$ is momentum measured by the SVD and the CDC), and
shower shape in the ECL. For weak electron identification, either
$dE/dx$ or $E/p$ is required to be consistent with the electron hypothesis.

\begin{figure}[htbp]
\begin{center}
\includegraphics[scale=.45,angle=0]{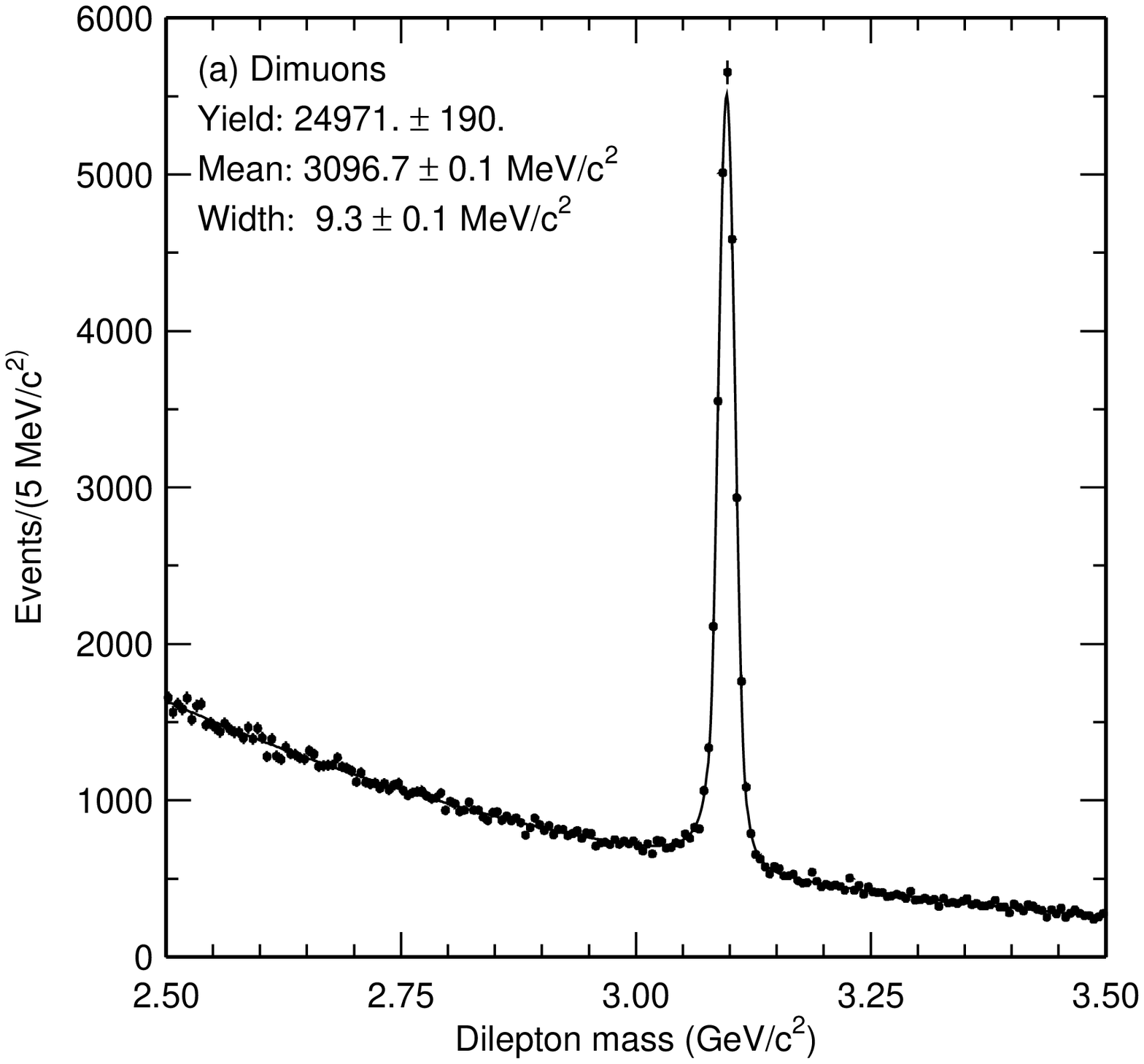} 
\includegraphics[scale=.45,angle=0]{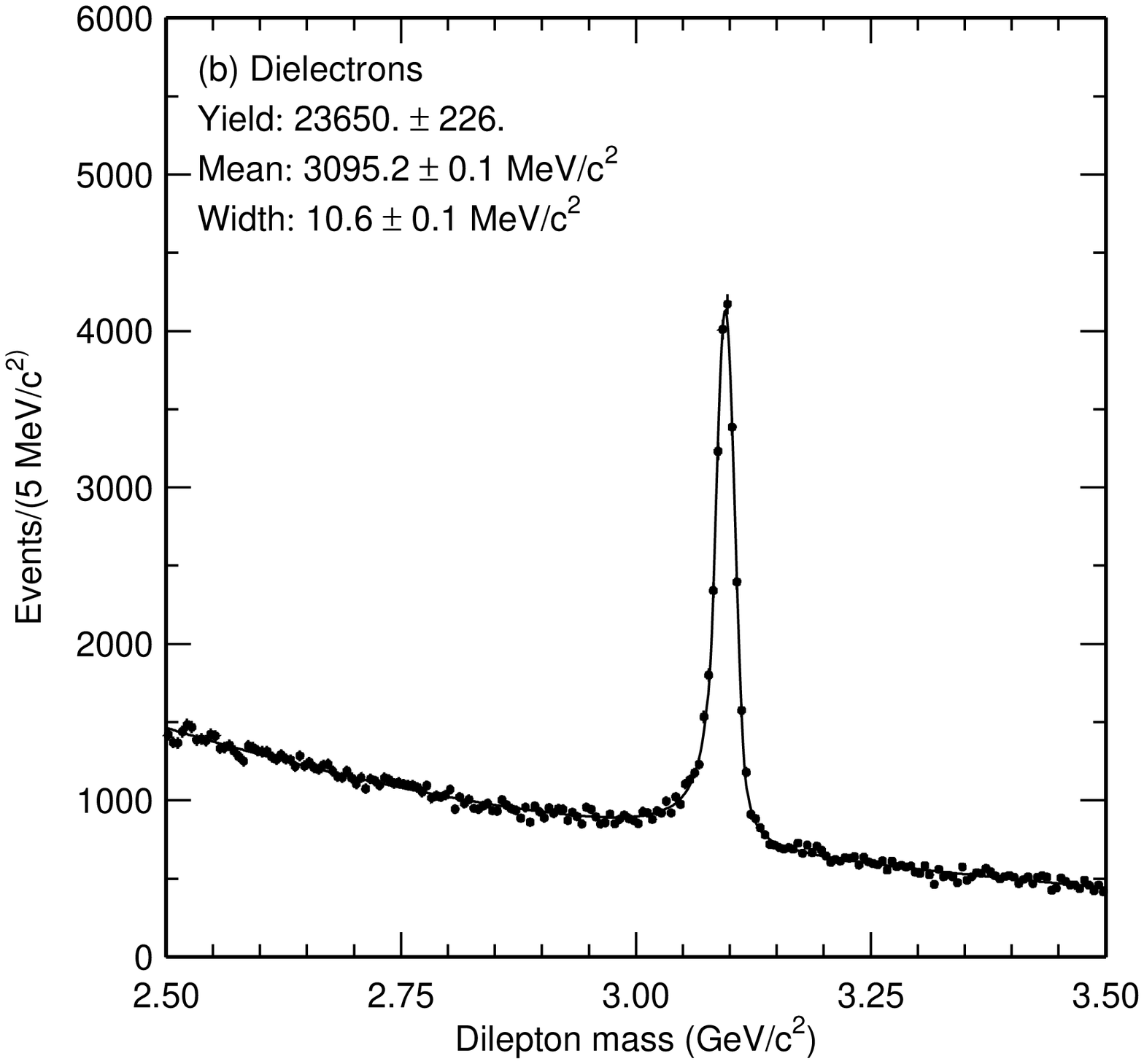}
\end{center} 
 \caption{
The invariant mass distributions for (a) $\mu^+\mu^-$ and 
(b)$e^+e^-$. In these figures, both leptons are tightly identified.
}
 \label{Jpsi-mass}
\end{figure}

For the identification of $J/\psi$ dilepton decays in the $B\to J/\psi K$
modes we require one tightly and one loosely identified lepton. For the
$\psi{\rm(2S)}K$ and $J/\psi \pi$ modes, both lepton candidates are
required to be tightly identified. We correct for final state
radiation or bremsstrahlung in the inner parts of the detector by
including the four-momentum of every photon detected within 0.05
radians of the original electron or positron direction in the
$e^+e^-$ invariant mass calculation. Figure~\ref{Jpsi-mass} shows the
dilepton mass distribution near the $J/\psi$ mass.  The mass
resolutions are 9.3 MeV/$c^2$ and 10.6 MeV/$c^2$ in the peak region
for $\mu^+\mu^-$ and $e^+e^-$, respectively.
Since there are still small radiative tails, as can be seen in
Figure~\ref{Jpsi-mass},
we use asymmetric invariant mass requirements, 
$( -60 < m_{\mu^+\mu^-} - m_{J/\psi(\psi {\rm(2S)})} < 36)\ {\rm MeV/}c^2$ and 
$(-150 < m_{  e^+  e^-} - m_{J/\psi(\psi {\rm(2S)})} < 36)\ {\rm MeV/}c^2$,
for the $\mu^+\mu^-$ and $e^+e^-$ pairs respectively.

\begin{figure}[htb]
 \begin{center}
\includegraphics[scale=.45,angle=0]{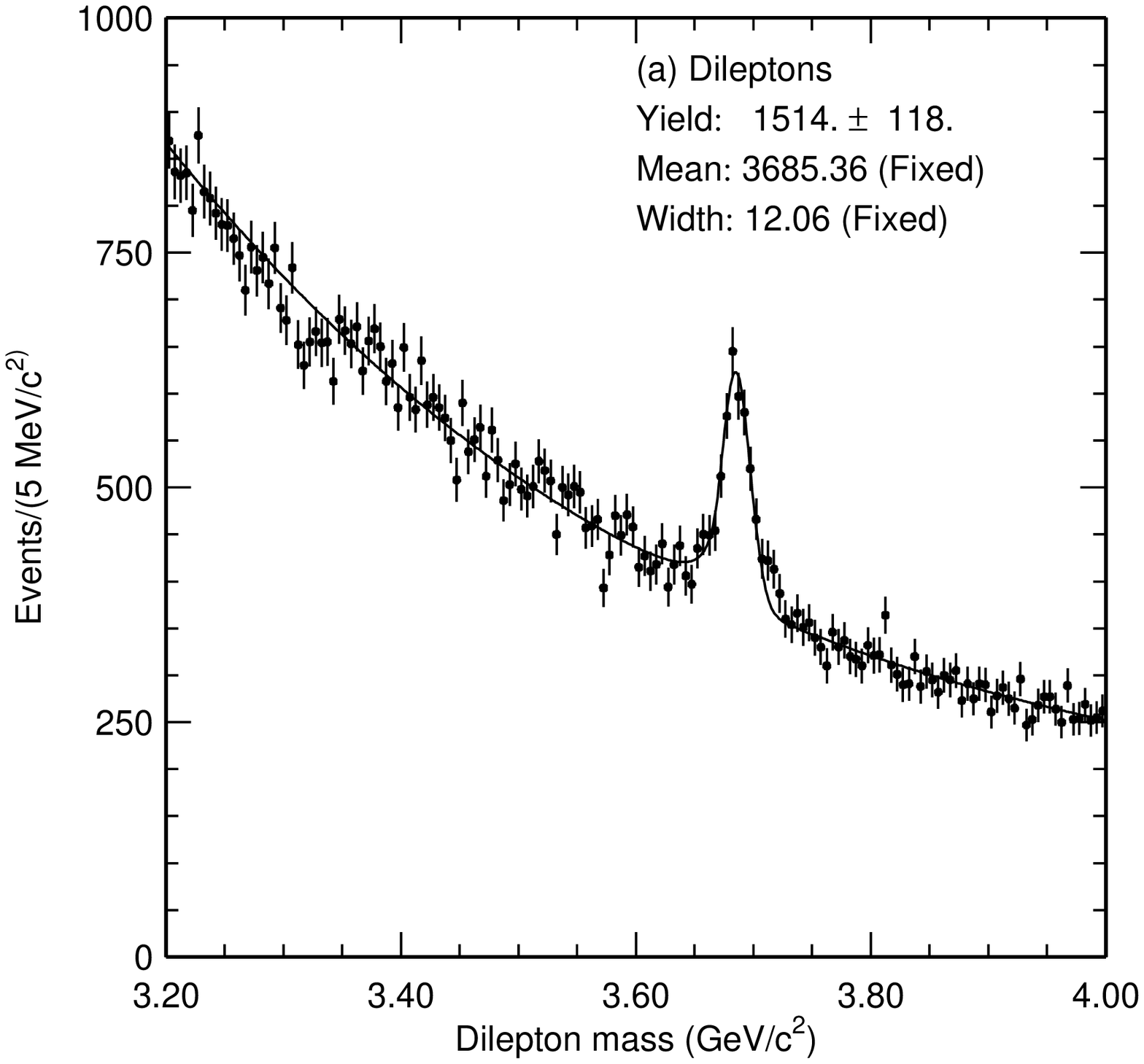} 
\includegraphics[scale=.45,angle=0]{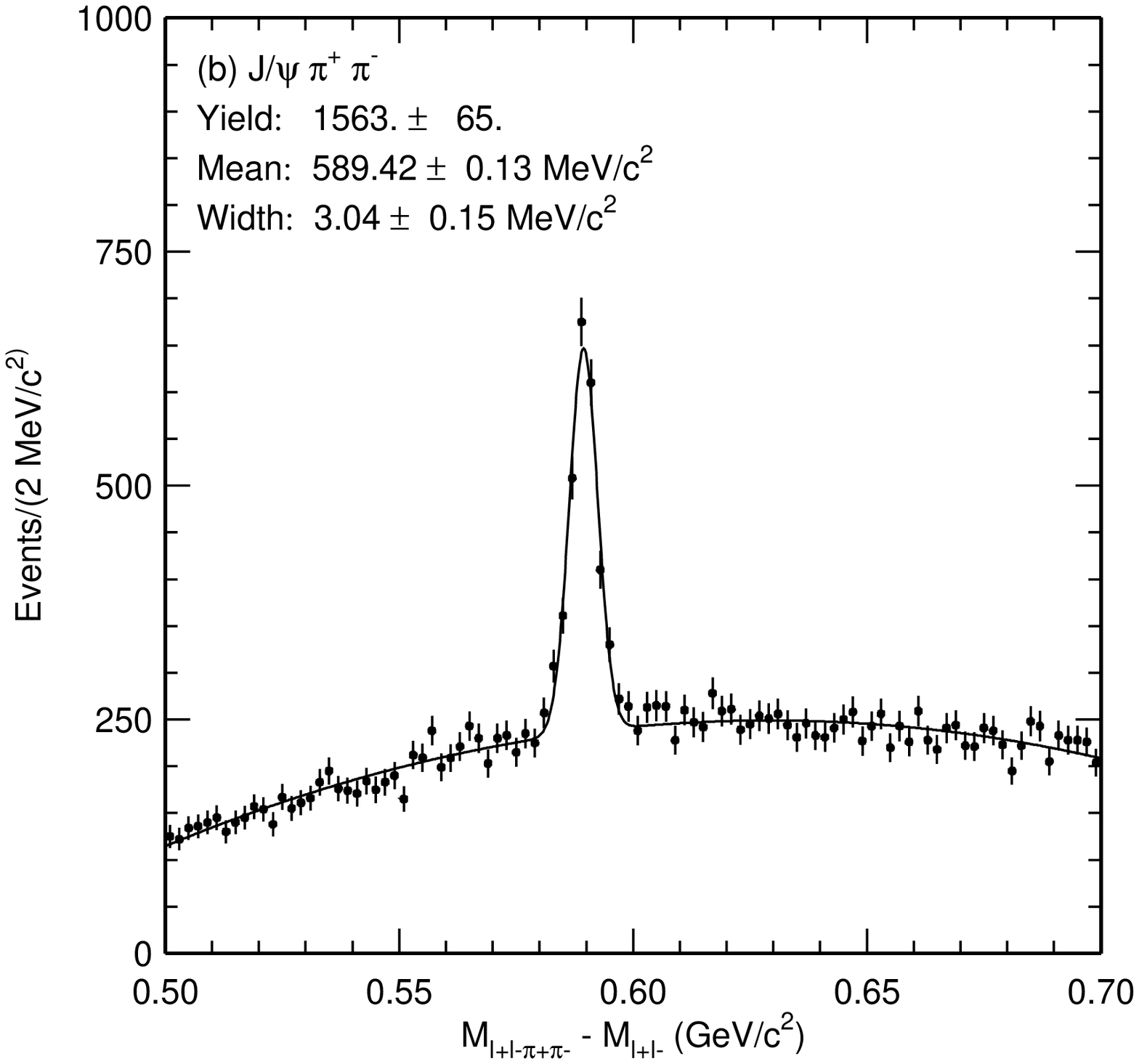} 
 \end{center}
\caption{(a) The invariant mass distribution of $\psi{\rm(2S)} \rightarrow l^+ l^-$
candidates, 
(b) the mass difference of $m_{l^+l^-\pi^+\pi^-}- m_{l^+l^-}$.}
\label{mass_of_psi2s}
\end{figure}

To identify $\psi{\rm(2S)} \rightarrow J/\psi\pi^+\pi^-$ candidates, 
we combine $J/\psi$ candidates with pairs of oppositely charged 
tracks that have a $\pi^+\pi^-$ invariant mass greater than 400 MeV/$c^2$. 
The $\psi$(2S) and $J/\psi$ candidates' mass difference is required to be
consistent with the known difference,
$(0.58<m_{l^+l^-\pi^+\pi^-}-m_{l^+l^-}<0.60)$~GeV/$c^2$. 
This range corresponds to $\pm 3\sigma$ in detector resolution.
Figure~\ref{mass_of_psi2s} shows (a) the invariant mass distribution 
of  $\psi{\rm(2S)} \rightarrow l l$, and 
(b) the mass difference of $m_{ll\pi\pi}- m_{ll}$.

\subsection{Light meson candidates}

In the analysis for $B^- \rightarrow J/\psi(\psi{\rm(2S)})K^-$,
all charged tracks (other than those used for $\psi$ reconstruction)
are used as kaon candidates in order to eliminate the systematic error
from particle identification.
This does not introduce any serious additional background,
because the principal background is expected to be from 
$B^- \rightarrow J/\psi(\psi{\rm(2S)})\pi^-$ decays, 
which occur at a much lower rate than $J/\psi(\psi{\rm(2S)})K^-$.
The prompt charged pion candidates are conversely required to be
strongly identified as pions ($P(\pi/K)>0.9$), 
where the likelihood ratio for a particle to be a charged pion, 
$P(\pi/K) = Prob(\pi)/(Prob(\pi)+Prob(K))$, is calculated 
using $dE/dx$ measured in the CDC and the response of the ACC.

For the analysis of neutral $B$ meson decays, the reconstruction of
$K^0_S \rightarrow \pi^+\pi^-$ is made by selecting pairs of
oppositely charged tracks with $\pi^+\pi^-$ invariant mass between 482
and 514~MeV/$c^2$.
This criterion retains 99.7\% of $K^0_S\rightarrow \pi^+\pi^-$ decays
with detected tracks, based on a double Gaussian
fit to the mass peak of the data (the average mass resolution is
4.4~MeV/$c^2$.).  In order to reduce combinatorial background further,
we require that:
\begin{itemize}
\item  
if both pions have associated SVD hits, 
the points of nearest approach of the two tracks 
in the projection onto the plane
perpendicular to the beam line ($r$-$\phi$) are separated in the beam
direction ($z$) by less than 1~cm,
\item
if only one of the two pions has associated SVD hits, 
the distance of nearest approach to the interaction point in the 
$r$-$\phi$ projection be
greater than 250~$\mu m$ for both tracks,
\item
if neither of the two pions have associated SVD hits,
the $\phi$ coordinate of the $\pi^+\pi^-$ vertex point and the $\phi$
direction of the $\pi^+\pi^-$ candidate's three-momentum agree 
within 0.1 radians.
\end{itemize}
The $K^0_S$ finding efficiency after track selection is 95\%.

In the selection of $B^0 \rightarrow J/\psi\pi^0$, the high momentum
$\pi^0$'s are reconstructed from pairs of detected photons.
The invariant mass is required to be 118
MeV/$c^2 < m_{\gamma\gamma} < $ 150 MeV/$c^2$ (mass resolution is
5.3~MeV/$c^2$). The $\pi^0$ candidate is also to have a good
mass constrained fit.

\subsection{$B$ Meson Reconstruction}

$B$ mesons are reconstructed by combining a charmonium meson candidate
with a kaon or pion candidate, as described above. The energy
difference, $\Delta E\equiv E_{\rm cand}-E_{\rm beam}$, and the beam-energy
constrained mass, $M_{\rm bc}\equiv \sqrt{E_{\rm beam}^2-P_{\rm cand}^2}$, are
used to separate signal from background ($E_{\rm beam}$ is the beam
energy, $E_{\rm cand}$ and $P_{\rm cand}$ are the $B$ candidate energy and
momentum, all calculated in the $\Upsilon$(4S) center of mass frame).

\begin{figure}
\begin{center}
\begin{tabular}{cc}
\includegraphics[scale=.45,angle=0]{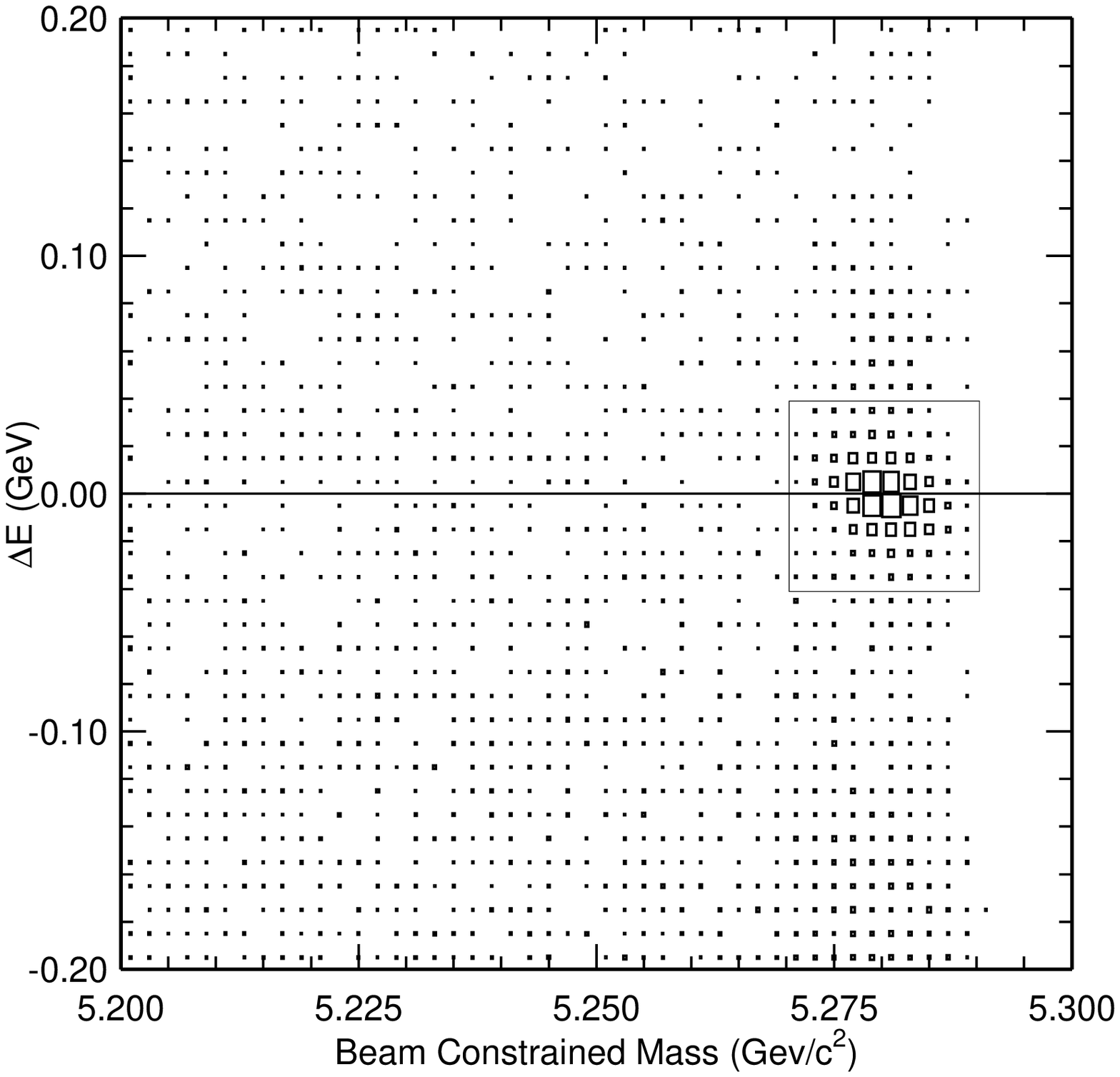} &
\includegraphics[scale=.45,angle=0]{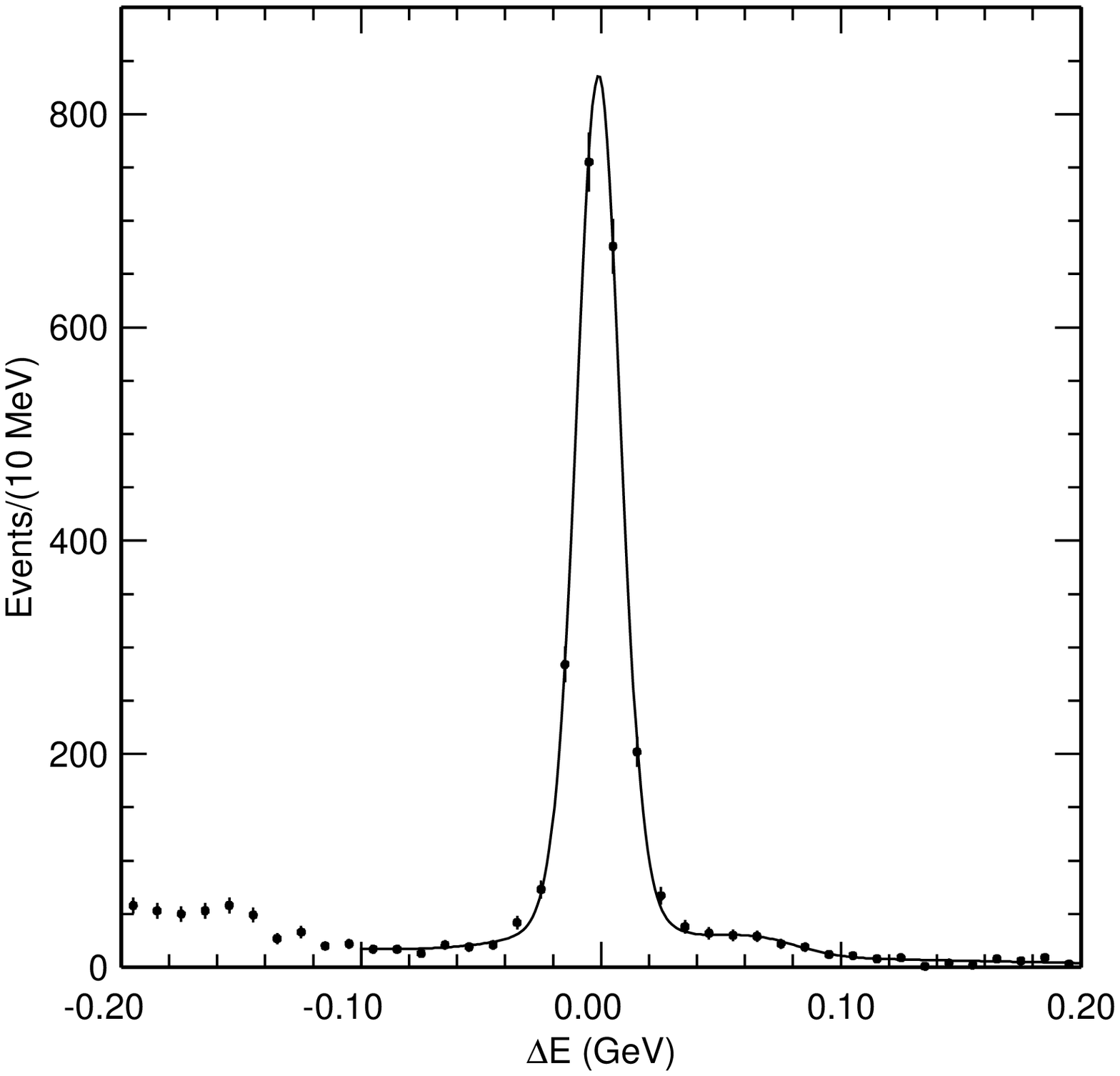} \\
\end{tabular}
\end{center}
\caption{The distribution of (a) $\Delta E$ versus $M_{\rm bc}$ and (b) $\Delta E$
for $B \rightarrow J/\psi K^{\pm}$.
The background from $B \rightarrow J/\psi K^*$ is seen at lower $\Delta E$,
while that from $B \rightarrow J/\psi \pi^{\pm}$ is at higher $\Delta E$.}  
\label{B-Jpsi}
\end{figure}

In this calculation, kinematic fits are performed with (1) mass and
vertex constraints for the $J/\psi$ or $\psi$(2S) di-lepton decays and
$K^0_S$ decays, and (2) a mass constraint for the $\psi{\rm(2S)} \to
J/\psi\pi^+\pi^-$ and $\pi^0 \to \gamma\gamma$ decays.
Figure~\ref{B-Jpsi} shows the distribution for $B\to J/\psi K^{\pm}$ candidates
in the $M_{\rm bc}$--$\Delta E$ plane as well as in $\Delta E$ after projecting
out candidates with $M_{\rm bc}$ between 5.27 and 5.29 GeV/$c^2$.
In the first plot an excess of candidates is clearly apparent in the signal
region, indicated by the rectangle.

In order to determine yields, we fit the $M_{\rm bc}$ distributions
after applying the following requirements on $\Delta E$: for all modes
except $J/\psi\pi$, $(-40 < \Delta E < 40)$ MeV;
for the $J/\psi \pi^0$ decay mode, $(-100 < \Delta E < 50)$~MeV, as the
$\Delta E$ distribution has a long tail at negative values due to
material in the detector and energy leakage;
for the $J/\psi \pi^-$ mode, $(-10 < \Delta E < 40)$~MeV, to suppress a
background from $B\to J/\psi K^-$ due to misidentification
of $K^-$ as $\pi^-$.

\begin{figure}
\begin{center}
\includegraphics[scale=0.60,angle=0]{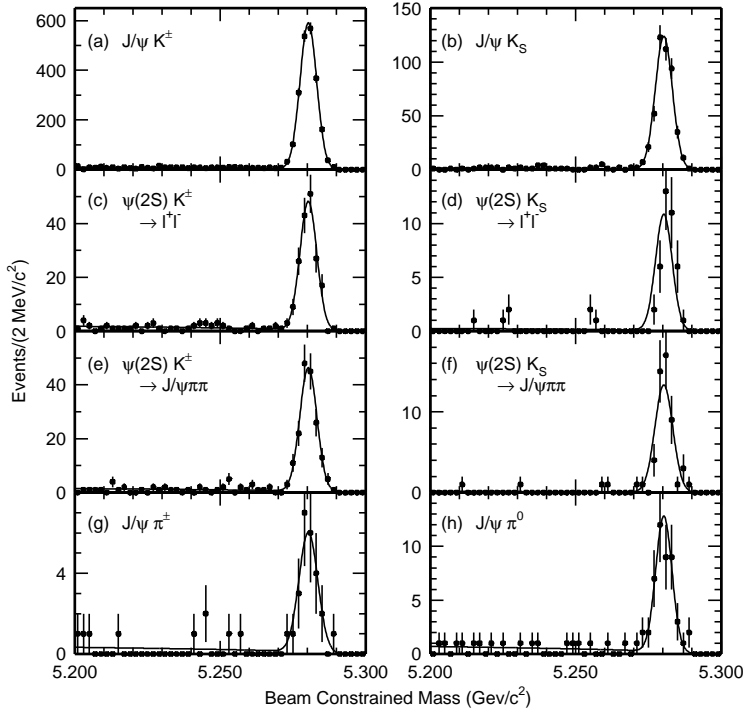} 
\end{center}
\caption{The distribution of $M_{\rm bc}$ for
(a) $B^-       \rightarrow J/\psi K^-$,
(b) $\bar{B^0} \rightarrow J/\psi K_S^0$, 
(c) $B^-       \rightarrow \psi{\rm(2S)} K^-  \{\psi{\rm(2S)} \rightarrow l^+l^-\}$,
(d) $\bar{B^0} \rightarrow \psi{\rm(2S)} K_S^0\{\psi{\rm(2S)} \rightarrow l^+l^-\}$,
(e) $B^-       \rightarrow \psi{\rm(2S)} K^-  \{\psi{\rm(2S)} \rightarrow J/\psi \pi^+ \pi^-\}$,
(f) $\bar{B^0} \rightarrow \psi{\rm(2S)} K_S^0\{\psi{\rm(2S)} \rightarrow J/\psi \pi^+ \pi^-\}$,
(g) $B^-       \rightarrow J/\psi \pi^-$ and
(h) $\bar{B^0} \rightarrow J/\psi \pi^0$.
}
\label{Mbc-all}
\end{figure}

The fit of the $M_{\rm bc}$ distribution is performed with the sum of a
Gaussian for signal and the ARGUS function\cite{argus} for
background (Figure~\ref{Mbc-all}). The resolution in $M_{\rm bc}$ is
dominated by the energy spread of KEKB. We test the resolution agreement
between MC and data using the mode $B^- \to J/\psi K^-$ (which has the
highest statistics). The agreement is very good and we use the MC predicted
widths for each mode to fit the $M_{\rm bc}$ histograms.
The signal yield and the
normalization of the background are allowed to vary in the fit.  The
results are shown in Table~\ref{exclusive_sum}.

\begin{figure}[htbm]
\begin{center}
\includegraphics[scale=0.5,angle=0]{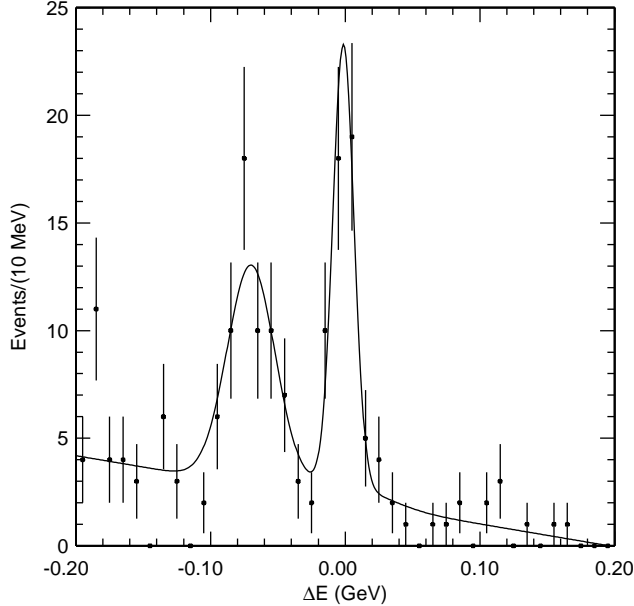} 
\end{center}
\caption{The $\Delta E$ distribution for $B^{\pm} \rightarrow
J/\psi\pi^{\pm}$. The signal peak is seen around zero. The peak
at $-0.07$~GeV/$c^2$ is from $B^{\pm} \rightarrow J/\psi K^{\pm}$.
In this figure, we require 5.27 $< M_{\rm bc} <$ 5.29 GeV/$c^2$}
\label{B-Jpsipi-dE}
\end{figure}

For the $J/\psi \pi^-$ mode, as shown in Figure~\ref{B-Jpsipi-dE}, the
background from $B^- \to J/\psi K^-$ peaks in the signal region of
$M_{\rm bc}$ but accumulates near $\Delta E\sim -70$~MeV due to
kinematic differences from the signal mode. To insure that it does
not intrude into the signal in the $M_{\rm bc}$ fit, a fit is performed on
the $\Delta E$ distribution with two separated Gaussians and a
first-order Chebyshev polynomial function (Figure~\ref{B-Jpsipi-dE}).
The signal yield obtained from the $\Delta E$ distribution fit is
consistent with the $M_{\rm bc}$ fit. The background yield from $J/\psi K^-$
is also consistent with the expectation from the mis-identification
probability.

\section{Results}
\subsection{Branching Fractions}

The reconstruction efficiencies are determined by Monte Carlo
simulations (MC) based on GEANT\cite{geant} and are listed in
Table~\ref{exclusive_sum}.  The number of $B\bar{B}$ events is
measured to be (31.9~$\pm$~0.3)$\times 10^6$. In the calculation of
the branching fraction, the production rates of $B^+B^-$ and
$B^0\bar{B^0}$ pairs are assumed to be equal.  We use the secondary
branching fractions listed in Table~\ref{substate} \cite{pdg}.  The
resulting branching fractions for each reconstructed decay chain are
summarized in Table \ref{exclusive_sum}, where the first errors are
statistical and the second are systematic.
The measurement values for the two $\psi {\rm (2S)}$ modes of
$B\rightarrow \psi{\rm (2S)} K$ are consistent within their
errors and the combined results are also listed in the table (taking
into account correlated and uncorrelated errors).

The sources of systematic error are shown in Table \ref{systematic}.
The dominant uncertainty arise from the uncertainty in the 
tracking efficiency.

The tracking efficiency uncertainty is determinied to be 2\% per track
from a comparison of the yields for $\eta \rightarrow \pi^+ \pi^-
\pi^0$ and $\eta \rightarrow \gamma\gamma$ with MC expectations.
The pion tracks from $K^0_S$ decay are from a displaced vertex and
thus may have a larger systematic error. We include a 3.5\% per
track uncertainty for these tracks (see below).

The uncertainty in the $K^0_S$ selection efficiency is checked by
comparing yields for a sample of high momentum $K^0_S\rightarrow
\pi^+\pi^-$ decays before and after applying the $K^0_S$ selection
criteria.  The efficiency difference between data and the Monte Carlo
simulation is less than 1.0\%.

As one cross-check of $K^0_S$ reconstruction, we also estimate the
difference between data and non-$\Upsilon$(4S) MC directly, using the yield
ratio between $D^0 \rightarrow K^0_S \pi^+ \pi^-$ and $D^0 \rightarrow
K^- \pi^+$ with $D^0$'s from $D^* \rightarrow D^0\pi$ decay.  In this
case, $D^0$'s with momentum higher than 3.0 GeV/$c$ are selected.  The
difference of the ratio between data and MC is also smaller than 1\%,
where a large systematic error arises from the uncertainties of the
world averages for the branching fractions.

The high momentum $\pi^0$ efficiency is checked by taking the ratio
between $D^0 \rightarrow K^+ \pi^- \pi^0$ and $D^0 \rightarrow K^+
\pi^- $ with high momentum $D^0$'s.  $D^0$'s generated from $D^*$ decay
with a slow pion are selected.  We assign a 7\% uncertainty to the
$\pi^0$ efficiency.

The efficiency of lepton identification is checked by comparing the
$J/\psi$ yield with one lepton tightly identified against the yield
where both leptons are tightly identified.  We find that the
efficiencies for tightly identified electrons and muons are 96\% and
94\%, respectively. The systematic errors from lepton identification
are determined to be 2\% per tightly identified lepton. The error for
loosely identified leptons in negligable.

For the $B^- \rightarrow J/\psi \pi^-$ mode, the identification of
high momentum charged pions is studied by comparing $D^{*+} \rightarrow
D^0 \pi^+$, where $D^0 \rightarrow K^-\pi^+$, between data and MC. 
In this decay mode, the $D^0$ mass peak is reconstructed with small
background without any particle identification requirements. The
systematic uncertainty is determined by examining the difference
in yield before and after applying particle identification. We assign
a systematic uncertainty of 2\% to the pion identification efficiency.

We also study the systematic error arising from the background in the
fit of the $M_{\rm bc}$ distribution.  The ARGUS function represents the
$M_{\rm bc}$ distribution for the $\Delta E$ sidebands well.
However there may be background decay modes that
peak in the signal region. We checked this with inclusive $J/\psi$ MC
and find no evidence for peaking background.

\begin{table}[htbm]
\begin{ruledtabular}
\caption{\label{exclusive_sum}
Signal yields and branching fractions for each mode.
Signal yields are determined by fitting. The errors are statistical
(first error) and systematic (second), except for the combined $\psi$(2S) modes where the
total error is listed. Efficiencies for modes with $K^0$ mesons are
for reconstructing $B\rightarrow  \psi K^0_S$.
}
\begin{tabular}{l l c c c }
\multicolumn{2}{c}{Decay Mode}          & Yield & Efficiency(\%) &B. F. ($\times 10^{-4}$) \\ 
\hline 
$B^- \rightarrow J/\psi \pi^- $         &       & $43.9\pm 6.8$ & 33.3 & $0.38\pm 0.06\pm 0.03$\\
$\bar{B^0} \rightarrow J/\psi \pi^{0} $ &       & $24.0\pm 5.0$ & 27.2 & $0.23\pm 0.05\pm 0.02$\\
\hline
$B^- \rightarrow J/\psi K^-$            &       & $2102\pm  46$ & 55.3 & $10.1\pm 0.2 \pm 0.7$ \\
$\bar{B^0} \rightarrow J/\psi K^0$      &       & $ 453\pm  21$ & 30.5 & $ 7.9\pm 0.4 \pm 0.9$ \\
\hline
$B^- \rightarrow \psi{\rm(2S)}K^-$      &       &               &      & $ 6.9\pm 0.6$         \\
 & $\psi{\rm(2S)} \rightarrow l^+l^-$           & $ 173\pm  13$ & 51.6 & $ 7.3\pm 0.6 \pm 0.7$ \\
 & $\psi{\rm(2S)} \rightarrow J/\psi\pi^+\pi^-$ & $ 170\pm  13$ & 23.2 & $ 6.4\pm 0.5 \pm 0.8$ \\
$\bar{B^0} \rightarrow \psi{\rm(2S)} K^0$ &     &               &      & $ 6.7\pm 1.1$         \\
 & $\psi{\rm(2S)} \rightarrow l^+l^-$           & $38.5\pm 6.2$ & 27.5 & $ 6.1\pm 1.0 \pm 0.8$ \\
 & $\psi{\rm(2S)} \rightarrow J/\psi \pi^+ \pi^-$         
	                                        & $51.2\pm 7.2$ & 12.0 & $ 7.4\pm 1.0 \pm 1.3$ \\
\end{tabular}
\end{ruledtabular}
\end{table}

\begin{table}
\caption{\label{substate} Branching fractions used for secondary charmonium decays\cite{pdg}.}
\begin{ruledtabular}
\begin{tabular}{ll}
 Decay mode       & Branching fraction   \\ 
\hline
$J/\psi         \to e^+e^-$             & 0.0593 $\pm$ 0.0010   \\ 
$J/\psi         \to \mu^+\mu^-$         & 0.0588 $\pm$ 0.0010   \\ 
$\psi{\rm(2S)}  \to e^+e^-$             & 0.0073 $\pm$ 0.0004   \\ 
$\psi{\rm(2S)}  \to \mu^+\mu^-$         & 0.0070 $\pm$ 0.0009   \\ 
$\psi{\rm(2S)}  \to J/\psi \pi^+\pi^-$  & 0.305  $\pm$ 0.016    \\
\end{tabular}
\end{ruledtabular}
\end{table}

\begin{table}
\caption{\label{systematic}The dominant sources of systematic errors (in \%).}
\begin{ruledtabular}
\begin{tabular}{l c c c c c c c c c} 
Decay mode& Tracking   & Lepton         & Hadron         & $K^0_S$($\pi^0$)& Charmonium & Monte & Total     \\ 
          & Efficiency & ID             & ID             & Efficiency      & Branching  & Carlo &           \\
          &            & Efficiency     & Efficiency     &                 & Fractions  & Stats.&           \\
\hline
$B^- \rightarrow J/\psi \pi^- $         &  6.0 & 4.0 & 2.0 &    & 1.2       & 1.7   &  7.8 \\
$\bar{B^0} \rightarrow J/\psi \pi^0 $   &  4.0 & 4.0 &     &7.0 & 1.2       & 1.9   &  9.3 \\ 
\hline 
$B^- \rightarrow J/\psi K^- $           &  6.0 & 2.0 &     &    & 1.2       & 1.4   &  6.6 \\
$\bar{B^0} \rightarrow J/\psi K^0$      & 11.0 & 2.0 &     &1.0 & 1.2       & 1.8   & 11.4 \\ 
\hline
$B^- \rightarrow \psi{\rm(2S)} K^-$         \\
~~~~$\psi{\rm(2S)} \rightarrow l^+l^-$  &  6.0 & 4.0 &     &    & 5.0       & 1.4   &  8.9 \\
~~~~$\psi{\rm(2S)} \rightarrow J/\psi \pi^+ \pi^-$ 
                                        & 10.0 & 4.0 & 4.0 &    & 5.3       & 2.1   & 12.8 \\ 
$\bar{B^0} \rightarrow \psi{\rm(2S)} K^0 $  \\
~~~~~$\psi{\rm(2S)} \rightarrow l^+l^-$ & 11.0 & 4.0 &     &1.0 & 5.0       & 1.9   & 12.9 \\
~~~~~$\psi{\rm(2S)} \rightarrow J/\psi \pi^+ \pi^-$ 
                                        & 15.0 & 4.0 & 4.0 &1.0 & 5.3       & 2.9   & 17.1 \\ 
\end{tabular}
\end{ruledtabular}
\end{table}

\subsection{Charge Asymmetries}

The yields for positive and negative $B$ mesons decays 
are measured separately using the method described above.
The charge asymmetries, defined by
\begin{eqnarray}
A_{K(\pi)} =
\frac{Br(B^-\rightarrow {\rm charmonium} + K^-(\pi^-)) - Br(B^+\rightarrow {\rm charmonium} + K^+(\pi^+))}
     {Br(B^-\rightarrow {\rm charmonium} + K^-(\pi^-)) + Br(B^+\rightarrow {\rm charmonium} + K^+(\pi^+))},
\end{eqnarray} 
(see \cite{CA_CLEO})
are calculated assuming the same efficiencies for both charged decays. 
The results are shown in Table \ref{Asym_charge}.
The efficiency difference between positive and negative particles is
determined by using $3.96\times 10^5$ and $3.33\times 10^5$ events for $D^{\pm}
\rightarrow K^{\mp} \pi^{\pm} \pi^{\pm}$ and
$D^0 \rightarrow K^- \pi^+$/$\bar D^0 \rightarrow K^+ \pi^-$ decays, respectively.
We calculate the efficiency ratios
$\epsilon_{\pi^-}/\epsilon_{\pi^+} = 1.011 \pm 0.015$ and 
$\epsilon_{K^-}/\epsilon_{K^+} = 1.004 \pm 0.017$ using the
following formulas:

\begin{eqnarray}
\frac{\epsilon_{\pi^-}}{\epsilon_{\pi^+}} = 
\frac{N(D^-\rightarrow K^+\pi^-\pi^-)N(          D^0 \rightarrow K^-\pi^+)}
     {N(D^+\rightarrow K^-\pi^+\pi^+)N(\overline{D^0}\rightarrow K^+\pi^-)},
\end{eqnarray} 

\begin{eqnarray}
\frac{\epsilon_{K^-}}{\epsilon_{K^+}} = 
\frac{N(D^-\rightarrow K^+\pi^-\pi^-)N(          D^0 \rightarrow K^-\pi^+)^2}
     {N(D^+\rightarrow K^-\pi^+\pi^+)N(\overline{D^0}\rightarrow K^+\pi^-)^2}.
\end{eqnarray} 

No significant efficiency differences
are observed for either pion or kaon tracks.
Thus, we do not correct the central values but we do include
the error of the efficiency differences in the systematic errors.
  
Finally, we find the charge asymmetries $-0.023 \pm 0.164 \pm 0.015$ and 
$-0.042 \pm 0.020 \pm 0.017$ for the charmonium+$\pi$ mode and 
the charmonium+$K$ mode, respectively.
Our results are consistent with zero asymmetry and previous measurements\cite{CA_CLEO,CA_BaBar}. 

\begin{table}
\caption{\label{Asym_charge}Charge asymmetry for each mode. Errors are statistical only.}
\begin{ruledtabular}
\begin{tabular}{l c c  c }  
  Decay mode & Yield($-$) & Yield(+) & $A_{K(\pi)}$\\ 
\hline 
  $B^{\pm} \rightarrow J/\psi\pi^{\pm}$          & $  21\pm  5$ &  $  22\pm  5$ &  $-0.023\pm 0.164$ \\
\hline
  $B^{\pm} \rightarrow J/\psi K^{\pm}$           & $1024\pm 32$ &  $1078\pm 33$ &  $-0.026\pm 0.022$ \\
  $B^{\pm} \rightarrow \psi{\rm(2S)}(l^+l^-) K^{\pm}$
                                                 & $  79\pm  9$ &  $  93\pm 10$ &  $-0.081\pm 0.078$ \\
  $B^{\pm} \rightarrow \psi{\rm(2S)}(J/\psi\pi^+\pi^-)K^{\pm}$
                                                 & $  68\pm  8$ &  $ 102\pm 10$ &  $-0.200\pm 0.075$ \\
\hline
 Total ($B^{\pm} \rightarrow J/\psi(\psi{\rm(2S)})K^{\pm}$)
                                                 & $1171\pm 34$ &  $1273\pm 36$ &  $-0.042\pm 0.020$ \\
\end{tabular}
\end{ruledtabular}
\end{table}

\section{Conclusion}

We have reported measurement of $B$ meson branching fractions to 
two-body final states that include a $J/\psi$ or $\psi$(2S) meson 
and a $K^0_S$, $K^{\pm}$, $\pi^0$ or $\pi^{\pm}$.
A total of 31.9 million $B\bar{B}$ events accumulated at the $\Upsilon$(4S)
resonance are used for this analysis.
Our results are in good agreement with previous measurements\cite{CLEO,BaBar}.
Charge asymmetries are also measured and found to be consistent with zero.

\section*{Acknowledgments}

We wish to thank the KEKB accelerator group for the excellent
operation of the KEKB accelerator.
We acknowledge support from the Ministry of Education,
Culture, Sports, Science, and Technology of Japan
and the Japan Society for the Promotion of Science;
the Australian Research Council
and the Australian Department of Industry, Science and Resources;
the National Science Foundation of China under contract No.~10175071;
the Department of Science and Technology of India;
the BK21 program of the Ministry of Education of Korea
and the CHEP SRC program of the Korea Science and Engineering Foundation;
the Polish State Committee for Scientific Research
under contract No.~2P03B 17017;
the Ministry of Science and Technology of the Russian Federation;
the Ministry of Education, Science and Sport of the Republic of Slovenia;
the National Science Council and the Ministry of Education of Taiwan;
and the U.S.\ Department of Energy.

\end{document}